\documentstyle[graphicx,twocolumn,eqsecnum,aps]{revtex}
\begin{document}
\draft
\title{Capillary condensation for fluids in spherical cavities}
\author{Ignacio Urrutia$^{1,}$\cite{urrucic} and Leszek
Szybisz$^{1,2,}$\cite{szybco}}
\address{$^1$Laboratorio TANDAR, Departamento de F\'{\i}sica,
Comisi\'on Nacional de Energ\'{\i}a At\'omica,\\
Av. del Libertador 8250, RA-1429 Buenos Aires, Argentina}
\address{$^2$Departamento de F\'{\i}sica, Facultad de
Ciencias Exactas y Naturales,\\
Universidad de Buenos Aires,
Ciudad Universitaria, RA-1428 Buenos Aires, Argentina}
\date{\today}
\maketitle
\begin{abstract}
The capillary condensation for fluids into spherical nano-cavities
is analyzed within the frame of two theoretical approaches. One
description is based on a widely used simplified version of the
droplet model formulated for studying atomic nuclei. The other, is
a more elaborated calculation performed by applying a density
functional theory. The agreement between both models is examined
and it is shown that a small correction to the simple fluid model
improves the predictions. A connection to results previously
obtained for planar slits and cylindrical pores is done.

FILENAME: cavity1.tex

\end{abstract}

\section{Introduction}
\label{sec:introduce}

Since more than a decade a so called simple fluid model (SFM) is
being applied for analyzing wetting properties of classical and
quantum fluids adsorbed onto planar substrates \onlinecite{chen,%
chiz}. The chief idea of this description is to write down the free
energy of the adsorbate as a sum of dominant volume, surface, and
substrate terms. As shown in Refs. \onlinecite{gat0} and
\onlinecite{calb}, the use of this SFM for studying capillary
condensation in the case of adsorption between two parallel planar
walls (i.e. in a slit geometry) yields a rich pattern of phase
transitions in agreement with more elaborated calculations carried
out with density functional (DF) approaches. More recently, the
SFM has been extended to investigate systems with cylindrical
symmetry \cite{cyan,less,gat1,sur,urs}. The authors of Ref.\
\cite{gat1} have found that for a rather compact regular array of
infinitely long, solid, and parallel cylinders, besides the well
known film and capillary condensation (CC) phases, an additional
necking configuration may appear. On the other hand, in Ref.\
\cite{sur} we have applied this SFM to describe CC transitions of
noble gases confined into single cylindrical pores of alkali metals
finding a nearly universal behavior. A detailed comparison of SFM
and DF results performed in Ref.\ \cite{urs} indicates a fair
agreement for the main features of cylindrical pores. The
phenomenon of hysteresis has been also addressed in Ref.\ \cite{%
sur}, it was found a good correspondence with results reported in
the pioneering work of Cole and Saam \cite{coal0}.

Currently there is an increasing interest for investigating the
storage capacity of small pores with different geometries (e.g.,
nanotubes and rather spherical zeolites, see Ref.\ \cite{gat1}
and references quoted therein). According to this trend, the
study of adsorption in spherical nano-cavities becomes relevant. 
Therefore the aim of the present work is to analyze phase
transitions for classical and quantum fluids of noble gases
confined into alkali metal cavities. We selected these systems
because the involved interactions are known. Its behavior has
been obtained from the SFM outlined above. In order to check the
validity of such a rather crude model for this geometry, the
adsorption of $^4$He is also calculated by applying a DF. The
latter approach has been recently successfully utilized for
investigating properties of $^4$He adsorbed on very small spherical
balls (with radii of a few \AA) of different materials \cite{sus,%
sur1}. However, for the sake of completeness, we should mention
that Monte Carlo \cite{bojan99,curta00,ancil01} as well as DF
\cite{ancil01,ancil99} calculations have been performed for
classical gases adsorbed on planar alkali metals.

The paper is organized in the following way. In Sec.\ \ref{%
sec:model} we describe the evaluation of the adsorption potential
and outline both the SFM and the DF approaches for capillary
condensation. The numerical results and an improvement of the bare
SFM are presented in Sec.\ \ref{sec:study} together with a
discussion of the hysteresis cycles and a summary of the whole
picture exhibited by the systems. Section \ref{sec:conclude} is
devoted to final remarks.

\section{The model}
\label{sec:model}

The properties of a fluid adsorbed by a solid substrate may be
studied by analyzing the grand free energy
\begin{equation}
\Omega[N] = F[N] - \mu\,N \:, \label{omega00}
\end{equation}
where $F$ is the Helmholtz free energy, $\mu$ the chemical
potential, and $N$ the number of particles of the adsorbate.
Quantity $F[N]$ contains the energy due to the interaction
between fluid atoms as well as the energy provided by the
confining potential. Let us first outline our approach for the
description of the fluid-substrate interaction.

\subsection{Adsorption potential}
\label{sec:potential}

In order to construct the physisorption potential inside a
spherical cavity we assume that an adsorbed atom located at
$\vec{r}$, measured from the pore center, interacts with an
elementary substrate volume at $\vec{r}\prime$ via an isotropic
(12,6) Lennard-Jones (LJ) pair potential with standard parameters
$\varepsilon_{LJ}$ and $\sigma_{LJ}$. Next, we suppose that
substrate atoms are uniformly distributed over the volume outside
the cavity, this continuous substrate approximation leads to a
spherically averaged potential. Under these conditions, when the
substrate extends from $r=R_p$ up to $r=\infty$, the total
potential may be expressed as
\begin{eqnarray}
U&&_{\rm sub}(r) = 8\pi\,\varepsilon_{LJ}\,\rho_{\rm sub}\,
\int^{\infty}_{R_p} r\prime^2 dr\prime \nonumber\\ 
&& \times \int^\pi_0\,\biggr[ \left(\frac{\sigma_{LJ}} {\mid
\vec{r} - \vec{r}\prime \mid} \right)^{12} - \left(\frac{
\sigma_{LJ}}{\mid \vec{r} - \vec{r}\prime \mid} \right)^6 \,
\biggr] \sin\theta\,d\theta  \:. \label{sub0}
\end{eqnarray}
Here, $\rho_{\rm sub}$ is the averaged density of the substrate in
number of atoms per unit volume. Of course, each approximation
performed along this procedure introduces an error. However, we
expect that the resulting potential would give a fair description
of the main features of examined systems. Notice that this
procedure has been previously adopted in Refs.\ \cite{hill,%
ricca} for getting the plain (9,3) potential in the case of planar
surfaces.

After some straightforward algebra the integral over $\theta$ may
be cast into the form
\begin{eqnarray}
I(r,r\prime)~&&= \int^\pi_0 \,\biggr[ \left(\frac{\sigma_{LJ}}{\mid
\vec{r} - \vec{r}\prime \mid} \right)^{12} - \left(\frac{
\sigma_{LJ}}{\mid \vec{r} - \vec{r}\prime \mid} \right)^6 \,
\biggr] \sin\theta\,d\theta \nonumber\\
&&= \left(\frac{\sigma_{LJ}}{r\prime} \right)^{12} M_{6}(\nu)
- \left(\frac{\sigma_{LJ}}{r\prime}\right)^6 M_3(\nu) \:,
\label{sub01}
\end{eqnarray}
where $M_n(\nu)$ stands for the integral
\begin{eqnarray}
M_n&&(\nu) = \int^\pi_0 \,\frac{\sin{\theta}\,d\theta}
{(1\,+\,\nu^2\,-\,2\,\nu\,\cos{\theta})^n} \nonumber\\
&&= \int^1_{-1} \,\frac{d x}
{(1\,+\,\nu^2\,+\,2\,\nu\,x)^n}
\nonumber\\
&&= \frac{1}{2(n-1)\nu} \biggr[ \frac{1}{(1-\nu)^{2(n-1)}}
-  \frac{1}{(1+\nu)^{2(n-1)}} \biggr] \;, \label{ems}
\end{eqnarray}
with $\nu=r/r\prime$. The integration over $r\prime$ yields the 
following expression for the adsorption potential
\begin{eqnarray}
U_{\rm sub}&&(r) =\frac{16\,\pi}{3}\,\varepsilon_{LJ}\,\rho_{\rm
sub}\,{R_p}^3 \nonumber\\ 
 && \times \biggr[ \frac{\left( 15\,r^6 + 63\,r^4\,{R_p}^2 + 
       45\,r^2\,{R_p}^4 + 5\,{R_p}^6 \right) \,\sigma^{12}_{LJ}}
     {15\,\left( R_p - r \right)^9\,\left( R_p + r \right)^9}
\nonumber\\
 &&~~~~ - \frac{\sigma^6_{LJ}}
     {\left( R_p - r \right)^3\,\left( R_p + r \right)^3} \biggr]
\:. \label{sub1}
\end{eqnarray}
This formula may be compared with expressions derived in Ref.\ 
\cite{which}. On the other hand, in the limit of a very large
cavity ($R_p \to \infty$) an expansion in terms of $z=R_p - r <<
R_p$ reduces Eq. (\ref{sub1}) to the (9-3) potential corresponding
to a planar substrate (cf. Eq. (4) in \cite{ricca})
\begin{equation}
U_{\rm sub}(z) = \frac{4\,\pi}{3}\,\varepsilon_{LJ}\,\sigma^3_{LJ}
\,\rho_{\rm sub}\,\biggr[\frac{1}{15}\left(\frac{\sigma_{LJ}}{z}
\right)^9 - \frac{1}{2}\left(\frac{\sigma_{LJ}}{z}\right)^3 \biggr]
\;. \label{plane}
\end{equation}

For all the inert gases, the substrate potentials produced by
pores in different alkali metals were calculated using values of
$\varepsilon_{LJ}$ and $\sigma_{LJ}$ determined by adjusting the
pair potentials of Patil \cite{pat} according to a procedure
outlined by Ancilotto {\it et al.} \cite{anil}. Density $\rho_{\rm
sub}$ was evaluated by taking into account that solid alkali metals
are bcc crystals with a lattice constants $a$ listed in Table 3 of
\cite{kit}. All the adopted parameters are quoted in Tables\
\ref{table1} and \ref{table2}.

\subsection{Simple fluid model}
\label{sec:simple}

An expression for the Helmholtz free energy per particle for the CC
phase, $f_{\rm CC}$, may be obtained from the formalism
developed in \cite{cyan,less}. Assuming that the free energy of the
fluid is given by the main contributions of the volume, surface,
and substrate terms; then, if the fluid fills a pore forming a
sharp sphere of radius $R_0$ with density equal to the bulk
equilibrium value $\rho_0$ one gets 
\begin{eqnarray}
f_{\rm CC} =&&~\frac{F_{\rm CC}}{N} = f_\infty + \sigma_{lv}\,
\left(\frac{36\,\pi}{\rho^2_0}\right)^{1/3}\,N^{-1/3} 
\nonumber\\
&&+ 4\,\pi\,\rho_0 \biggr\{\,\int^{R_0}_0 r^2\,dr\,U_{\rm sub}(r)
\biggr\}\,N^{-1} \;, \label{foot}
\end{eqnarray}
with
\begin{equation}
N = 4\,\pi \int^{R_0}_0 {r^2\;dr\;\rho(r)}
= \frac{4\,\pi}{3}\,\rho_0\,R^3_0 \;. \label{number}
\end{equation}
Here $f_\infty$ is the asymptotic value and it coincides with the
chemical potential $\mu_0$ at saturated vapor pressure (SVP).
This expression for the free energy has been recently utilized for
studying adsorption into cylindrical pores \cite{less,sur}.

In all the cases, for a spherical cavity with a certain radius
$R_p$ the potential $U_{\rm sub}(r)$ given by Eq.\ (\ref{sub1})
exhibits a minimum with depth $D_{R_p}$ at a distance $r_m(R_p)$
from the wall. It can be analytically shown that for a given
fluid-substrate combination, in the regime $R_p > \sigma_{LJ}$,
the depth $D_{R_p}$ decreases for increasing size of the spherical
cavity and attains an asymptotic value $D_\infty$ according to the
expression
\begin{eqnarray}
&& D_{R_p} = D_{\infty}\,\biggr[ 1 
         + \frac{27\,R^*}{4\,{\left( 2\,R^* - 1\right) }^2} 
         - \frac{3}{8\,{\left( 2\,R^* - 1 \right) }^3}
\nonumber\\
&& - \frac{5 - 54\,R^* + 216\,{R^*}^2 - 504\,{R^*}^3 
  + 576\,{R^*}^4 - 288\,{R^*}^5}{8\,{\left( 2\,R^*- 1 \right)}^9}
\biggr] \:, \nonumber\\
\label{depth0}
\end{eqnarray}
where
\begin{equation}
D_\infty = \frac{2 \sqrt{10}}{9}\,\pi\,\rho_{\rm sub}\,
\varepsilon_{LJ}\,\sigma_{LJ}^3 \:, \label{dinfty}
\end{equation}
and being the dimensionless pore radius
\begin{equation}
R^* = R_p/r_m \:. \label{ar}
\end{equation}
Here $r_m$ is the asymptotic position of the minimum [$r_m(R_p
\to \infty)$] which is given by
\begin{equation}
r_m = \left( \frac{2}{5} \right)^{1/6} \sigma_{LJ} \:. \label{rm}
\end{equation}

The behavior of $D_{R_p}/D_\infty$ as a function of $1/(R^*-1/2)$
is similar to that previously obtained for the cylindrical
geometry (see Figs. 1 and 2 in \cite{sur} and Fig. 1 in \cite{%
stucco} ). In order to facilitate any comparison the values of
$r_m$ and $D_\infty$ are listed in Table\ \ref{table2}. Since
the properties found for the adsorption into cavities in Rb and
K do not differ much from those for pores in Cs, the corresponding
values are not included in this table. Note that for a given
fluid-substrate system, the present $r_m$ is larger than the
corresponding parameter for a slit geometry $z_m$ determined from
a (9,3) potential in Ref.\ \cite{gat0} (cf. Table I therein). This
feature is simply due to different approaches used to fix the
potential parameters.

\subsubsection{Capillary condensed droplet phase}
\label{sec:capful}

Upon taking into account Eqs.\ (\ref{foot}) and (\ref{number}),
and defining the sharp sphere radius $R_0$ as the effective pore
radius $R_0=R_p-r_m$, the grand thermodynamic potential for the
CC phase becomes
\begin{eqnarray}
\Omega_{\rm CC}~&&= F_{\rm CC}-\mu\,N \nonumber\\
&&= 4\,\pi\,\sigma_{lv}\,\left(R_p-r_m\right)^2
+ 4\,\pi\,\rho_0 \int^{R_p-r_m}_{0} r^2\,dr\,U_{\rm sub}(r)
\nonumber\\
&&~~~-  \frac{4\,\pi}{3}\,(\mu-\mu_0)\,\rho_0\,
\left(R_p-r_m\right)^3 \:. \label{omega1}
\end{eqnarray}
Starting from this equation, it is possible to get the modified
Kelvin equation (KE) for CC. By setting $\Omega_{\rm CC}=0$, one
arrives at
\begin{equation}
\sigma_{sl} - \frac{1}{3}\,(\mu-\mu_0)\,\rho_0\,\left(R_p -
r_m\right) = 0 \:, \label{kelvin0}
\end{equation}
where $\sigma_{sl}$ is the liquid-substrate interfacial tension
defined as the sum
\begin{eqnarray}
\sigma_{sl}~&&= \sigma_{lv} + \frac{\rho_0}{(R_p-r_m)^2}
\int^{R_p-r_m}_{0} r^2\,dr\,U_{\rm sub}(r) \nonumber\\
&&= \sigma_{lv} + I_U \:, \label{sigma}
\end{eqnarray}
Next, by using the Gibbs-Duhem \cite{row} relation for the fluid
and neglecting any compression of the system one gets a link to
the pressure $P$
\begin{equation}
\rho_0\,(\mu-\mu_0) = P - P_0 \:,
\label{relay}
\end{equation}
which leads to the usual KE
\begin{equation}
3\,\sigma_{sl} - (P - P_0)\,\left(R_p-r_m\right) = 0 \:.
\label{kelvin}
\end{equation}
This equation expresses the pressure reduction for condensation in
terms of the effective radius of the condensed fluid $(R_p-r_m)$.
The purpose of this work is to generalize the KE by taking into
account both: (i) the explicit dependence of $\Omega_{\rm CC}$ on
the substrate potential and (ii) the role of film formation on the
adsorption behavior.

For the sake of generality, we shall discuss the predictions in
terms of the dimensionless grand potential
\begin{equation}
\Omega^* = \frac{\Omega}{A\,\sigma_{lv}} \:, \label{omega0}
\end{equation}
which for the CC phase may be written as
\begin{eqnarray}
\Omega^*_{\rm CC}~&&= \frac{\Omega_{\rm CC}}{A\,\sigma_{lv}}
= \frac{F_{\rm CC}-\mu\,N}{4\,\pi\,(R_p-r_m)^2\,\sigma_{lv}}
\nonumber\\
&&= 1 + \frac{I_U}{\sigma_{lv}} -\frac{1}{3}\,\frac{(\mu-\mu_0)
\,\rho_0}{\sigma_{lv}}\,\left(R_p-r_m\right) \:. \label{omegas}
\end{eqnarray}
It is convenient to cast the integrated adsorption potential per
unit area $I_U$ in the following reduced form
\begin{eqnarray}
I^*_U~&&= \frac{I_U}{\sigma_{lv}} \nonumber\\
&&= - \frac{D_\infty\,\rho_0}{\sigma_{lv}(R_p-r_m)^2}
\int^{R_p-r_m}_{0} r^2\,dr\,[-U_{\rm sub}(r)/D_\infty] \nonumber\\
&&= - \frac{1}{2}\,D^*\,g\{R^*-1\} \;, \label{gee1}
\end{eqnarray}
Here, the strength parameter is the reduced asymptotic well depth
\begin{equation}
D^* = 2\,\frac{D_\infty\,r_m\,\rho _0}{\sigma_{lv}}
= \frac{4\,(20)^{1/3}\,\pi}{9}\,\rho_{\rm sub}\,\varepsilon_{LJ}\,
\sigma_{LJ}^4\,\frac{\rho_0}{\sigma _{lv}} \:. \label{Dee}
\end{equation}
and $g\{R^*-1\}$ is the dimensionless integral
\begin{equation}
g\{\xi\} = \frac{1}{\xi^2} \int^\xi_0 \xi\prime^2\,d\xi\prime\,
[-U_{\rm sub}(r_m\,\xi\prime)/D_\infty] \:, \label{gee}
\end{equation}
with
\begin{equation}
\xi = r/r_m \:. \label{wy}
\end{equation}
Consequently, one has also to introduce the reduced difference of
chemical potentials
\begin{equation}
\Delta = (\mu_0-\mu)\,r_m\,\rho_0/\sigma_{lv} \:. \label{Del}
\end{equation}
These definitions of $D^*$ and $\Delta$ are formally equal to those
adopted in previous works \cite{gat0,sur}, provided that for planar
parallel walls instead of $r_m$ one uses $z_m$.

Finally, one may write the reduced version of the grand potential
given by Eq.\ (\ref{omegas}) in the following way
\begin{equation}
\Omega^*_{\rm CC} = 1 - \frac{1}{2}\,D^*\,g\{R^*-1\} + \frac{1}{3}
\,\Delta\,\left(R^*-1\right) \:. \label{omega2}
\end{equation}
This expression allows a study of adsorption in terms of $D^*$.
Before going ahead, let us remind that a fundamental property of
the integral $g\{\xi\}$ introduced by Eq.\ (4) in \cite{gat0} is
its independence of the fluid-substrate combination. This feature
is also exhibited by the integral $g\{\xi\}$ calculated according
to Eq.\ (\ref{gee})
\begin{eqnarray}
g\{\,\xi\,\} =~&&-\frac{2}{3}\,\frac{{R^*}^3\,\xi\,( 5\,{R^*}^4
+ 14\,{R^*}^2\,\xi^2 + 5\,\xi^4 )}{({R^*}^2 - \xi^2)^8}
\nonumber\\
&&+ \frac{3}{2}\,\frac{R^*\,({R^*}^2 + \xi^2)}
{\xi\,({R^*}^2 - \xi^2)^2} - \frac{3}{4}\,\frac{1}{\xi^2}\,
\ln \biggr[\frac{R^* + \xi}{R^* - \xi}\biggr] \;. \label{gee2}
\end{eqnarray}
For $\xi=R^*-1$, in the limiting case $R^* \to \infty$ one gets
\begin{equation}
\lim_{R^*\to\infty}\,g\{R^*-1\} = \frac{11}{16} \;. \label{gee3}
\end{equation}
This result is equal to that obtained for the slit geometry
\cite{gat0}.

\subsubsection{Shell-film phase}
\label{sec:film}

Under the same assumptions adopted for writing Eqs.\ (\ref{foot})
and (\ref{omega1}), the grand thermodynamic potential for a
spherical film of thickness $\ell$, i.e., for the shell-film (SF)
phase, may be expressed as
\begin{eqnarray}
\Omega_{\rm SF}~&&= F_{\rm SF}-\mu\,N \nonumber\\
&&= 4\,\pi\,\sigma_{lv}\,[\left(R_p-r_m\right)^2
+ \left(R_p - r_m - \ell \right)^2] \nonumber\\
&&~~~+ 4\,\pi\,\rho_0 \int^{R_p-r_m}_{R_p-r_m-\ell} r^2\,dr\,
U_{\rm sub}(r) \nonumber\\
&&~~~- \frac{4\,\pi}{3}\,(\mu-\mu_0)\rho_0[\left(R_p-r_m\right)^3
- \left(R_p - r_m - \ell\right)^3] \:. \nonumber\\
\label{omega11}
\end{eqnarray}
Note that the film grows from $r=r_m$ towards the center. The
reduced version of this grand free energy reads
\begin{eqnarray}
\Omega^*_{\rm SF}~&&= \frac{\Omega_{\rm SF}}{4\,\pi\,R_0^2\,
\sigma_{lv}} = 1 + \left(1 - \frac{x}{R^*-1}\right)^2 - \frac{1}{2}
\,D^* \nonumber\\
&&~~~\times \biggr[\,g\{R^*-1\} - \left(\,1 - \frac{x}{R^*-1}\,
\right)^2 g\{R^*-1-x\}\,\biggr] \nonumber\\
&&~~+ \frac{1}{3}\,\Delta\,\left(R^*-1\right)\,\biggr[\,1
- \left(\,1 - \frac{x}{R^*-1}\,\right)^3\,\biggr] \:,
\label{omega12}
\end{eqnarray}
where $x$ is the dimensionless thickness
\begin{equation}
x = \ell/r_m \:, \label{exec}
\end{equation}
which is related with the dimensionless inner radius $y$ measured
from the center of the sphere
\begin{equation}
y = (R_p-r_m-\ell)/r_m = R^*-1-x \:. \label{inner}
\end{equation}
In this case, in order to find the stable configuration at fixed
$R^*$ (i.e., $R_p=const.$), one must determine the value of $x$
which provides the minimum $\Omega^*_{\rm SF}$.

\subsection{Density functional theory}
\label{sec:functional}

In the DF approach the ground-state energy of an interacting
$N$-body system of $^4$He atoms, confined by an adsorbate-substrate
potential $U_{\rm sub} ({\bf r})$, may be written as
\begin{eqnarray}
E_{\rm gs}~&&= -{\hbar^2\over2 m} \int { d{\bf r} \,\sqrt{\rho({\bf
r})}\,{\bf \nabla}^2 \sqrt{\rho({\bf r})}} + \int { d{\bf r}\,
\rho({\bf r})\,e_{sc}({\bf r})} \nonumber\\
&&~~~+ \int { d{\bf r}\,\rho({\bf r})\,U_{\rm sub}({\bf r}) } \;,
\label{Ene0}
\end{eqnarray}
where $\rho({\bf r})$ is the one-body density and $e_{sc}({\bf r})$
the self-correlation energy per particle. The density profile
$\rho({\bf r})$ is determined from the Euler-Lagrange (EL) equation
derived from the condition
\begin{equation}
\frac{\delta \Omega}{\delta \rho({\bf r})}
= \frac{\delta \{\,E_{\rm gs}[\rho,{\bf \nabla}\rho] - \mu\,N\,\}}
{\delta \rho({\bf r})} = 0 \;. \label{vary0}
\end{equation}
In the case of a spherical symmetry the variation of Eq.\
(\ref{vary0}) leads to the following Hartree like equation for the
square root of the one-body helium density
\begin{eqnarray}
-\frac{\hbar^2}{2 m} \:&& \left(\,\frac{d^2}{dr^2}
+ \frac{2}{r}\,\frac{d}{dr} \right) \,\sqrt{\rho(r)} \nonumber\\
&&+~\biggr[\,V_H(r) + U_{\rm sub}(r) \biggr] \,\sqrt{\rho(r)}
= \mu \, \,\sqrt{\rho(r)} \;, \label{hairs}
\end{eqnarray}
which also determines $\mu$. Here $V_H({\bf r})$ is a Hartree
mean-field potential given by the first functional derivative of
the total correlation energy $E_{sc}[\rho]$
\begin{equation}
V_H({\bf r}) = \frac{\delta E_{sc}[\rho]}{\delta \rho({\bf r})}
= \frac{\delta}{\delta \rho({\bf r})} \, \int d{\bf r}\prime \,
\rho({\bf r}\prime)\,e_{sc}({\bf r}\prime) \;. \label{harp}
\end{equation}
The expression for the spherically symmetric $V_H( r)$ derived
in the case of the Orsay-Paris nonlocal DF (OP-NLDF) proposed in
\cite{dhpt} is given in the Appendix of \onlinecite{sur2}.
Equation (\ref{hairs}) was solved at a fixed number of helium
atoms
\begin{equation}
N = 4\,\pi \int^\infty_0 {r^2\;dr\;\rho(r)} \;.
\label{number2}
\end{equation}

\section{General Results - Phase Diagrams}
\label{sec:study}

\subsection{Thresholds for CC and SF phases at SVP}
\label{sec:thresh0}

\begin{figure}
\centering\includegraphics[width=6cm, angle=-90]{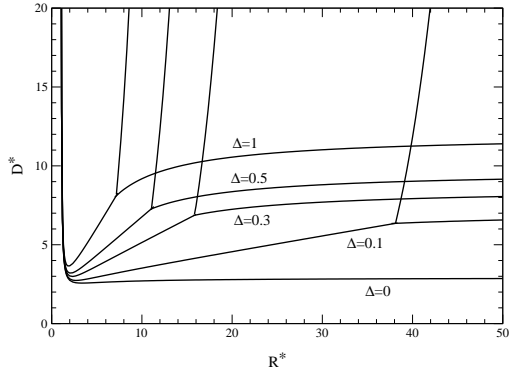}
\caption{Phase diagram showing the behavior at various
degrees of undersaturation, expressed in terms of the reduced
chemical potential difference $\Delta$ given by Eq.\ 
(\protect\ref{Del}). For the case $\Delta=0$, the line satisfies
Eq.\ (\protect\ref{dee0}); all values above the curve correspond
to capillary condensation, while those below are ``empty''. For
the other cases, $\Delta > 0$, there is a ``triple point'' below
which the space is empty, to the upper right of which there is a
shell film, and to the upper left there is capillary condensation.}
\label{phase1}
\end{figure}

The case of SVP ($\Delta=0$) leads to a very simple criterion for
the occurrence of CC. The expression for the transition line
($\Omega^*_{\rm CC} = 0$) which separates the behavior into two
regimes, capillary condensation ($\Omega^*_{\rm CC} < 0$) or empty
($\Omega^*_{\rm CC} > 0$), is
\begin{equation}
D^*_{\Delta=0}({\rm CC}) = \frac{2}{g\{R^*-1\}} \:. \label{dee0}
\end{equation}
It coincides with Eq.\ (7) of \cite{gat0} and with Eq.\ (3.1) of
\cite{sur}. Its dependence on $R^*$ is displayed in Fig.\
\ref{phase1}. Therefore, one can state that the curve given by Eq.\
(\ref{dee0}) provides a universal relation for the critical values
of the parameters $R^*$ and $D^*_{\Delta=0}$. On the other hand, it
should be noticed that at SVP the transition from empty (E) to SF
would occur for
\begin{eqnarray}
&&D^*_{\Delta=0}({\rm SF}) \nonumber\\
&& = \frac{2\,(R^*-1)^2+2\,(R^*-1-x)^2}
{(R^*-1)^2g\{R^*-1\}-(R^*-1-x)^2g\{R^*-1-x\}} \:. \nonumber\\
\label{dee1}
\end{eqnarray}
It is clear that, for $0<x<R^*-1$, $D^*_{\Delta=0}({\rm SF})$ is
always larger than $D^*_{\Delta=0}({\rm CC})$. So, for increasing
$D^*$ at SVP, before $\Omega^*_{\rm SF}$ becomes zero
$\Omega^*_{\rm CC}$ is already negative favoring the CC phase
against the SF one.

Furthermore, the threshold condition for CC to occur at SVP for
large $R^*$ changes very little for increasing $R^*$ and for
very broad pores attains the asymptotic value
\begin{equation}
D^*_{\Delta=0}({\rm CC};R^*\to\infty) = \lim_{R^*\to\infty}\,
\biggr[ \frac{2}{g\{R^*-1\}}\biggr] = \frac{32}{11} \simeq 2.9 
\:. \label{dee2}
\end{equation}
A glance at Table\ \ref{table2} indicates that noble gases heavier
than $^4$He adsorbed in cavities of Cs would not form in the
present approach a stable CC phase at the triple-point temperature
$T=T_t$, because the corresponding values of $D^*$ are smaller
than 2.9.

\subsection{Thresholds for CC and SF phases below SVP}
\label{sec:thresh1}

The general problem of behavior below SVP is more complicated. The
presence of the $\Delta$ term in Eq.\ (\ref{omega2}) leads to the
following threshold value for CC
\begin{equation}
D^*({\rm CC}) = D^*_{\Delta=0}({\rm CC})\,\biggr[\,1 + \frac{1}
{3}\,\Delta\,(R^*-1)\,\biggr] \:. \label{dee3}
\end{equation}
However, one must now examine the possibility that a film
configuration has a lower free energy than that of the CC phase.
So, in the case $\Delta > 0$, it is necessary to evaluate the
minimum of $\Omega^*_{\rm SF}$ as a function of the variable $x$
and to compare the result with $\Omega^*_{\rm CC}$.

The reduced depth for E $\to$ SF transition at $\Delta > 0$ may
be obtained from Eq.\ (\ref{omega12}), written in terms of $y$ it
becomes
\begin{equation}
D^*({\rm SF}) = D^*_{\Delta=0}({\rm SF})\,\biggr[\,1
+ \frac{1}{3}\,\Delta\,\frac{ (R^*-1)^3 - y^3}{(R^*-1)^2 + y^2}
\,\biggr] \:. \label{dee4}
\end{equation}

The difference between the grand free energy of the SF phase and
that of the CC case written in terms of the dimensionless inner
radius $y$ is
\begin{equation}
\Omega^*_{\rm SF} - \Omega^*_{\rm CC} = \left( \frac{y}{R^*-1}
\right)^2 \left(\,1 + \frac{1}{2}\,D^*\,g\{y\} - \frac{1}{3}\,
\Delta\,y\,\right) \:. \label{omega5}
\end{equation}
Here, in the second parenthesis, the first term represents the
extra surface energy of the film, the second provides the
interaction between the solid substrate and the adsorbed atoms
which fill the gap when the SF to CC transition occurs; while the
third stands for the free energy cost because the system is (in
general) below SVP. When the internal radius of the shell film
goes to zero, i.e. $y \to 0$, the difference of grand potentials
given by Eq.\ (\ref{omega5}) vanishes. The threshold value for
the transition SF $\to$ CC is
\begin{equation}
D^*({\rm SF \to CC}) = \frac{2}{g\{y\}}\,\biggr[\,-1 
+ \frac{1}{3}\,\Delta\,y\,\biggr] \:, \label{dee5}
\end{equation}
provided that $y$ is taken at the minimum of $\Omega^*_{\rm
SF}$.

\begin{figure}
\centering\includegraphics[width=6cm, angle=-90]{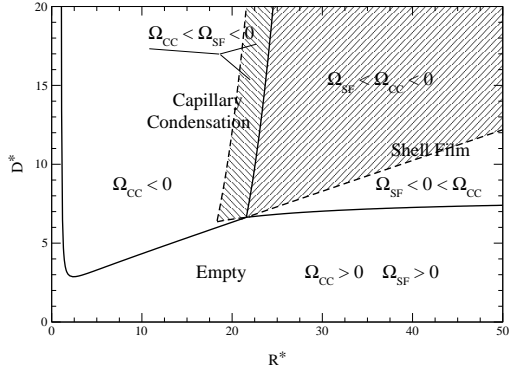}
\caption{Universal ``phase diagram'' showing regimes of empty,
capillary condensed, and adsorbed shell film as a function of
reduced pore radius and well depth defined in the text [see Eqs.\
(\protect\ref{ar}) and (\protect\ref{Dee})]. The displayed curves
correspond to $\Delta=0.2$. The hatched zones are regions
where both $\Omega^*_{\rm SF}$ and $\Omega^*_{\rm CC}$ are
negative.}
\label{phase0}
\end{figure}

Before describing the general phase diagram of the systems for
any $\Delta > 0$, it is useful to examine with some detail
data calculated for a fixed value of $\Delta$. Figure \ref{phase0}
shows results obtained for $\Delta = 0.2$ by varying $D^*$ and
$R^*$. In this case some features of $\Omega^*_{\rm CC}$ and
$\Omega^*_{\rm SF}$ are explicitly indicated. Solid lines
separate domains of stable phases. Note that there is a ``triple
point'' below which the space is empty (E), to the upper right of
which there is a SF region, and to the upper left there is CC.
The dashed straight line in the SF region is the threshold given
by Eq.\ (\ref{dee3}), while dashed curves in the CC regime denote
the limit for SF solutions determined by the threshold given by
Eq.\ (\ref{dee4}) and the disappearance of the minimum of
$\Omega^*_{\rm SF}$. So, the hatched zones indicate parameter
regions where both CC and SF are negative. However, only that
phase with lower $\Omega^*$ is stable, the other one is metastable
and plays an important role in the cycle of hysteresis to be
addressed later in the paper.

Let us now turn to the general phase diagram. The results for
several values of $\Delta$ are displayed in Fig.\ \ref{phase1},
where only boundaries between stable phases are indicated.
Increasing $D^*$ favors SF or CC phases against the E phase.
Which of the condensed phases is stable depends on $R^*$. For
large $R^*$, SF is typically favored because of the cost of CC
(the $\Delta$ term) becomes large relative to the benefits (from
the potential and the decrease of surface energy). The SF to E
transition curve is rather insensitive to $R^*$.

Figures \ref{phase2} and \ref{phase3} show the reduced phase
diagrams projected onto the $R^* - \Delta$ plane for $D^*=5.84$
and $11.87$ (i.e. for the $^4$He/Cs and $^4$He/Na cases,
respectively). However, for other values of $D^*$ the shape of the
corresponding curves is similar. The curves indicate the
transitions between stable phases. These plots also include
threshold values obtained from OP-NLDF calculations for $R_p=10$,
$12$, $15$, $20$, and $30$~\AA. For $^4$He/Na these results lie
in the neighbor of the ``triple point'' determined by the joining
of all three E, SF, and CC phases, where the solutions are
sensitive to changes of $\Delta$. Note that the E $\to$ CC and SF
$\to$ CC transitions provided by the SFM occur at almost the same
values of $\Delta$ as that determined from OP-NLDF. There is no SF
phase for $^4$He/Cs systems in the analyzed range of $R^*$. On the
other hand, the predictions for the E $\to$ SF transition do not
match so well. All these features of the agreement are similar to
that previously found for planar and cylindrical systems (see Fig.
4 in \cite{gat0} and Fig. 6 in \cite{sur}).

\subsection{Correction to the bare simple fluid model}
\label{sec:improve}

The prediction given by the bare SFM for the E $\to$ SF phase
transition may be improved by introducing a correction to the
surface term in $\Omega_{\rm SF}$. By looking at the surface energy
term in Eq.\ (\ref{omega11}), which is proportional to
$\sigma_{lv}$, one realizes that it does not vanish in the limit of
a zero-thickness film (i.e. for $\ell \to 0$). In order to
eliminate this shortcoming one may follow an idea adopted by Cheng
{\it et al.}\cite{chen0} in writing their Eq.\ (2.4). Accordingly,
we shall assume that the surface contribution grows exponentially
from zero at $\ell=0$ to the bare value $4\pi\sigma_{lv}[\left(R_p-
r_m\right)^2+\left(R_p -r_m-\ell\right)^2]$ over a characteristic
length $\zeta$
\begin{eqnarray}
\Omega_{\rm SF}({\rm surf})~&&= 4\,\pi\,\sigma_{lv}\,
[\left(R_p-r_m\right)^2 + \left(R_p - r_m - \ell \right)^2]
\nonumber\\
&&~~~\times \biggr[ 1 - \exp(-\ell/\zeta) \biggr] \:.
\label{omegas1}
\end{eqnarray}
This leads to the reduced version
\begin{equation}
\Omega^*_{\rm SF}({\rm surf}) =
\biggr[ 1 + \left(1 - \frac{x}{R^*-1}\right)^2 \biggr]
\biggr[ 1 - \exp\left(-\beta\,x\right) \biggr] \:, 
\label{omegas2}
\end{equation}
with
\begin{equation}
\beta = r_m/\zeta \:. \label{beta}
\end{equation}

Let us now analyze the changes of the energetics due to this
cut-off factor. As expected, its effect only becomes important for
rather thin films. In the frame of the improved SFM the E $\to$ SF
transition is reached for a larger value of $\mu_0-\mu$
than in the original version. Both these features can be observed
in Fig.\ \ref{cave0}, where results for $^4$He adsorbed into a
spherical Na cavity with radius $R_p=30$~\AA$\,$are shown. Hence,
the more elaborated approach yields a better agreement with OP-NLDF
results. In fact, for the $^4$He/Na systems a very good agreement
is obtained with $\beta_a=3.3$ as shown in Fig.\ \ref{phase3}. It
should be noted that the corrected curves for the E $\to$ CC and
SF $\to$ CC transitions cannot be distinguished from the previous
ones on the scale of the drawing.

It is plausible to assume that the characteristic length $\zeta$
be mainly proportional to the asymptotic width $W_\infty$ of a
free $^4$He surface. So, one may write
\begin{equation}
\beta = r_m/\zeta = \lambda\,r_m/W_\infty = \lambda/W^* \:,
\label{betas}
\end{equation}
where $W^*$ is the dimensionless asymptotic width
\begin{equation}
W^* = W_\infty/r_m \:. \label{width}
\end{equation}
By using the asymptotic value $W_\infty \simeq 6$~\AA$\,$\cite{%
sur2}, for $^4$He/Na systems one gets $W^* \simeq 1.3$, which leads
to $\lambda_a \simeq 4.3$. A critical film thickness $\ell_a$ may
be defined by requiring that the exponential of Eq. (\ref{omegas1})
be reduced from unity to about 0.1, i.e. so that the surface energy
reach a 90\% of its bare value of Eq.\ (\ref{omega11}). This choice
for the decay coincides with the density fall off adopted for
defining the surface thickness in the literature \cite{sur2}.
Following this procedure, the obtained value of $\lambda_a$ yields
$x_a/W^*=\ell_a/W_\infty \simeq 0.54$. This means that, in the
improved SFM the bare value of the surface energy will be
essentially reached when an adsorbed film of uniform density
$\rho_0$ would have approximately the number of helium atoms enough
to develop a surface with thickness $W_\infty$ at its internal
face.

In Figs.\ \ref{cave1} and \ref{cave2} we compare the chemical
potentials, the free energies and grand free energies per particle
provided by the improved SFM with results obtained from the OP-NLDF
calculations. The data correspond to $^4$He adsorbed into cavities
of Cs and Na with radius $R_p=30$~\AA. These plots show a good
qualitative agreement between both theoretical approaches, of
coarse, there are some quantitative differences. Note that in the
case of $^4$He/Na a Maxwell construction is required in order to
determine $\mu_0-\mu=3.7$~K from OP-NLDF data for the SF $\to$ CC
transition.

An example of the density profile evolution as a function of
the number of $^4$He atoms is shown in Fig.\ \ref{profiles}.
There OP-NLDF results, including the metastable (or unstable)
solutions, obtained for a Na cavity with radius $R_p=30$~\AA$\,$
are displayed. One may observe that the growth of the density
profile is continuous and relatively smooth throughout the
available space.

It is worthy of notice that the value $\beta_a=3.3$ fixed above
lies just in the parameter region determined by Cheng {\it et al.}
\cite{chen0} for approaching results given by the full NLDF theory
in the case of planar geometry. Moreover, this $\beta_a$ is close
to $\beta=3$ chosen in that paper for the purpose of illustration.
Therefore it becomes of interest to make a connection to the paper
of Calbi {\it et al.}\cite{calb} on $^4$He confined by two parallel
walls separated by a distance $L$. By looking at Figs.\ 15 and 16
of that work one may realize that predictions for the E $\to$ F
(film) phase transitions provided by the bare SFM do not reproduce
quite well results given by the NLDF theory. So, we examined the
extend to which a correction similar to that included in Eq.\
(\ref{omegas1}) can improve the agreement in the case of adsorption
into slits. For such a geometry, the surface contribution to the
reduced grand free energy is given by the factor 4 in Eq.\ (8) of
Ref.\ \cite{gat0}. Consequently, that term was rewritten as
\begin{equation}
\Omega^*_{\rm F}({\rm surf}) = 4\,\biggr[ 1 - \exp\left(-
\frac{\lambda\,x}{W^*}\right) \biggr] \:, \label{omegas3}
\end{equation}
here $W^*$ is
\begin{equation}
W^* = W_\infty/z_m \:. \label{widths}
\end{equation}
All the reduced dimensionless quantities needed to treat a planar
system can be obtained from Eqs.\ (1)-(6) in \cite{calb}. To
illustrate the effect of this cut-off factor we selected the case
of $^4$He adsorbed between planar walls of Li. The evaluation was
performed with the same cut-off parameter $\lambda_a \simeq 4.3$
utilized for spherical cavities. Figure\ \ref{phase4} shows data
reported in Fig.\ 15 of \cite{calb} together with the shift of the
E $\to$ F transition towards larger $\Delta$. From this plot one
may conclude, that also for the planar geometry the agreement with
NLDF results is significantly improved. As before the E $\to$ CC
and SF $\to$ CC transitions remain almost unchanged.

\subsection{Critical radii and adsorption process}
\label{sec:radii}

Let us now refer to the phenomenon of hysteresis inherently related
to capillary condensation. For the analysis to be made in this
section it is convenient to define a reduced inner radius $\eta$ as
\begin{equation}
\eta = \frac{y}{R^*-1} =1-\frac{x}{R^*-1} \;. \label{eta}
\end{equation}
Since the empty fraction of the cavity is $\eta ^3$, for $\eta=0$
capillary condensation with a completely filled pore occurs, while
in the $\eta \to 1$ limit for large $R^*$ there are very narrow
shell films adsorbed on the substrate wall. We determined two
families of critical inner radii: $\eta_M$ at which the solution
for SF becomes metastable and $\eta_C$ at which SF becomes
unstable. Associated to the critical radii $\eta_M$ and $\eta_C$
there are critical reduced chemical potentials $\Delta_M$ and
$\Delta_C$. Numerous experimental evidences reported in the
literature indicate that porous materials fill and drain at
different values of the chemical potential leading to loops of
hysteresis (see Lilly and Hallock\cite{hall1} and references quoted
therein). In order to explain these loops it was suggested by Cole
and Saam\cite{coal0} that the process of adsorption and desorption
of a single cavity do not always follow the path guided by stable
phases. The idea is that physical systems exhibit a sort of memory
trying to remain in the initial phase when $\Delta$ is changed.

The filling and draining of a spherical cavity may be followed by
looking at Fig. 6 of Ref.\ \onlinecite{hall1}. This process can be
analyzed by examining Figs. \ref{phase3} and \ref{coals}. For a
fixed pore size and a given adsorbate-substrate combination, the
adsorption starts at $\Delta>0$ in an E phase, which corresponds to
the vapor phase of the fluid. Next, the filling of the pore by
decreasing $\Delta$ increases the vapor density (this behavior is
not included in the SFM, in which all the gas states are considered
E phase). At some value of $\Delta$ the SF phase becomes stable and
a thin shell film is formed, then it grows increasing its thickness
$x$ and reducing the inner radius $\eta$ up to the crossing of
$\Omega^*_{\rm SF}$ with $\Omega^*_{\rm CC}$. The key assumption
for building an elementary hysteretic loop\cite{coal0} is that
after this crossing the filling follows the metastable SF phase up
to the critical point ($\Delta_C$, $x_C$) where it becomes
unstable. Then the SF $\to$ CC transition must occur as indicated
in Fig. 8(b) of \cite{sur}. By draining the transition CC $\to$ SF
occurs at ($\Delta_M, \eta_M$) giving rise to an elementary loop of
hysteresis. If $R^*$ is sufficiently small the SF phase cannot
exist and the filling jumps directly from E to CC following a
simpler path and no hysteresis takes place. In fact, the whole
process described here is equal to that assumed in the case of
cylindrical pores\cite{sur,coal0,hall1}.

Figure \ref{coals} shows results obtained for adsorption of noble
gases into cavities of Na and Li, for these adsorbate-substrate
systems $D^*$ is always larger than 2.9. For each system the {\it
nearly} universal behavior starts at about $R^*-1 \simeq 15$,
which for $D^*$ in the range $4 \to 20$ yields for the abscissa 
\begin{equation}
\Psi = \frac{R^*-1}{\sqrt{D^*}} \label{Psi}
\end{equation}
values between $7.5$ and $3.6$. The limiting curves for metastable
and unstable SF solutions may be derived from expansions in powers
of $1/\sqrt{D^*}$. Upon keeping the leading term only, the explicit
formula for the metastable border reads
\begin{equation}
\Psi^2_M = \frac{3}{8}\,\biggr[\,\frac{6\,\eta^4+16\,\eta^2-6}
{\eta\,(1-\eta^2)^3} + \frac{3}{\eta^2}
\,\ln \left(\frac{1 + \eta}{1 - \eta}\right) \biggr]  \:,
\label{Psi1}
\end{equation}
while for the unstable one becomes
\begin{equation}
\Psi^2_C = \frac{18\,\eta^3}{(1-\eta^2)^4} \:. \label{Psi2}
\end{equation}
These functions are also plotted in Fig.\ \ref{coals}, where one
may observe that its behavior is similar to that found before
for cylindrical pores in terms of hypergeometric functions (see
Fig.\ 10 in \cite{sur} and Fig.\ 2 in \cite{coal0}). The limiting
curve corresponding to E $\to$ SF lies at $\eta = 1$.

\section{Final remarks}
\label{sec:conclude}

A SFM is used for studying adsorption into isolated spherical
cavities. This model has previously been successfully applied for
analyzing phase transitions in the cases of planar slits
\cite{gat0} and cylindrical pores \cite{sur}. For the spherical
geometry treated in the present work the SFM yields a {\it
universal} description of capillary condensation transitions for
noble gases confined in pores of alkali metals providing an
interpretation of the wide range of behavior which can occur. This
{\it universality} is due to the fact that the integrated
adsorption potential $g\{\xi\}$ expressed in dimensionless
quantities [see Eqs.\ (\ref{gee}) and (\ref{gee2})] is a {\it
universal} function of $\xi=r/r_{\rm min}$. This property is a
consequence of the continuous substrate approximation. Planar slits
and cylindrical pores also exhibit this feature, but in the last
geometry to get $g\{\xi\}$ one must perform numerical integrations
\cite{sur}.

The phase diagram for spherical cavities displayed in Fig.\ \ref{%
phase1} is qualitatively equivalent to that previously obtained in
for planar and cylindrical geometries (see Fig. 2 in \cite{gat0}
and Fig. 3 in \cite{sur}).

\begin{figure}
\centering\includegraphics[width=6cm, angle=-90]{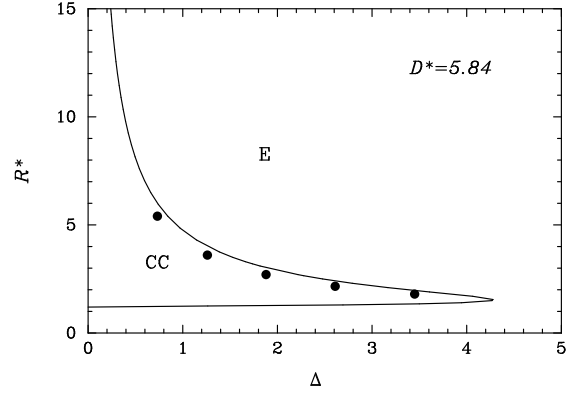}
\caption{Reduced phase diagram for $D^*=5.84$ (i.e. $^4$He/Cs in
our approach). The curve indicates the prediction of the SFM for
the boundary between possible stable phases (empty and capillary
condensation). The full circles stand for transitions determined
from OP-NLDF calculations.}
\label{phase2}
\end{figure}

\begin{figure}
\centering\includegraphics[width=6cm, angle=-90]{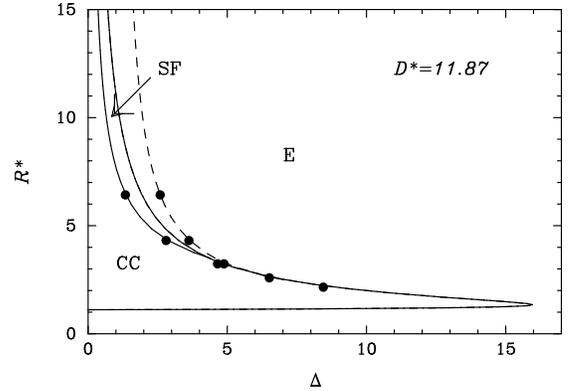}
\caption{Reduced phase diagram for $D^*=11.87$ (i.e. $^4$He/Na
in our approach). The solid curves indicate the prediction for the
boundaries among possible stable phases (empty, shell film, or
capillary condensation) according to the SFM. Note the existence of
a ``triple point''. The full circles stand for transitions
determined from OP-NLDF calculations. The dashed curve stand for
the E $\to$ SF transition given by the improved SFM.}
\label{phase3}
\end{figure}

The reliability of the SFM was checked by performing a comparison
with results provided by the much more realistic OP-NLDF. In the
latter case the calculations were carried out for increasing number
of $^4$He atoms within spherical cavities of different radii. It
was found that for the E $\to$ CC and SF $\to$ CC phase transitions
there is a rather good agreement between both approaches (see
Figs.\ \ref{phase2} and \ref{phase3}), while for the E $\to$ SF one
there is a sizeable difference as indicated by Fig.\ \ref{phase3}.

\begin{figure}
\centering\includegraphics[width=6cm, angle=-90]{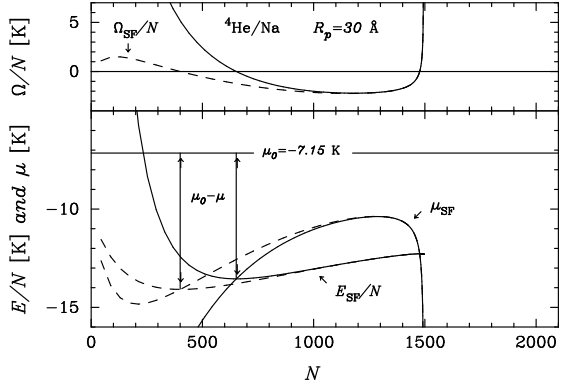}
\caption{Energetics of $^4$He adsorbed into a Na cavity with
$R_p =30$~\AA. {\it Upper part:} The grand free energy per
particle as a function of the number of particles. {\it Lower
part:} Same for the energy per particle and the chemical
potential. The solid curves are results of the original SFM while
dashed curves are provided by the improved version.}
\label{cave0}
\end{figure}

\begin{figure}
\centering\includegraphics[width=6cm, angle=-90]{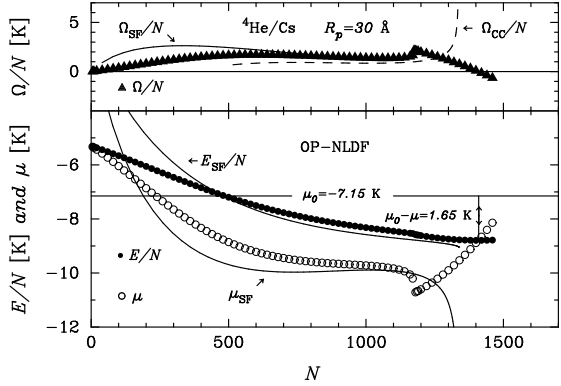}
\caption{Energetics of $^4$He adsorbed into a Cs cavity with
$R_p =30$~\AA. {\it Upper part:} The grand free energy per
particle as a function of the number of particles; {\it Lower
part:} Same for the energy per particle and the chemical
potential. The curves indicate results provided by the improved
SFM. The symbols stand for values obtained from OP-NLDF
calculations.}
\label{cave1}
\end{figure}

\begin{figure}
\centering\includegraphics[width=6cm, angle=-90]{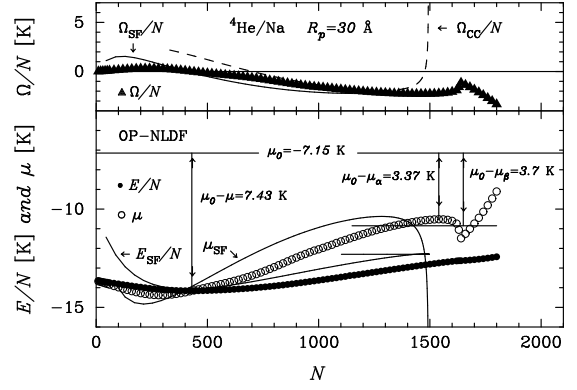}
\caption{Same as Fig.\ \protect\ref{cave1} for $^4$He adsorbed
into a Na cavity.}
\label{cave2}
\end{figure}

\begin{figure}
\centering\includegraphics[width=6cm, angle=-90]{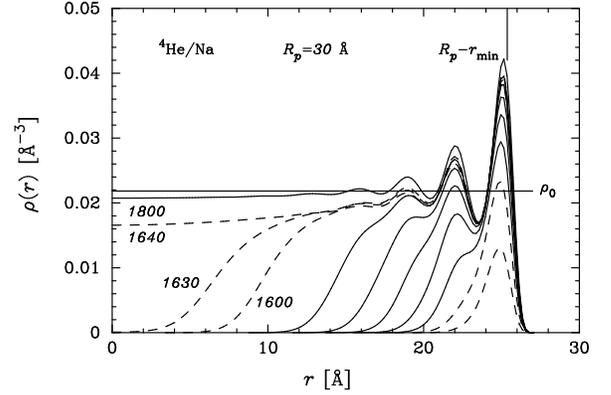}
\caption{Density profiles of $^4$He adsorbed into a cavity of
Na ($R_p=30$~\AA) for $N=200$, 400, 600, 800, 1000, 1200, 1400,
1600, 1630, 1640, and 1800. Dashed curves are metastable or
unstable solutions.}
\label{profiles}
\end{figure}

\begin{figure}
\centering\includegraphics[width=7cm]{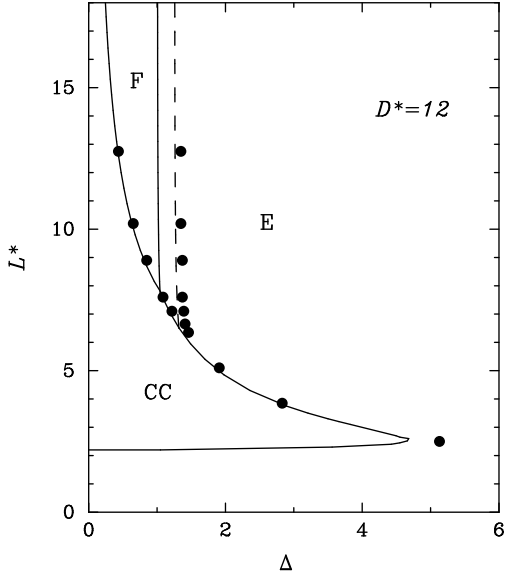}
\caption{Reduced phase diagram for $^4$He confined by two planar
walls of Li ($D^*=12$). Solid curves are SFM predictions and full
circles DF results, both of them taken from
\protect\onlinecite{calb} (see text). The dashed curve shows how
much the correction introduced in the present work improves the
prediction of the SFM for the E $\to$ F phase transition.}
\label{phase4}
\end{figure}

It is shown that the bare SFM for shell films given by Eq.\
(\ref{omega11}) may be improved by including a correction to the
surface contribution. As can be seen in Figs.\ \ref{phase3},
\ref{cave0}, and \ref{cave2}, the effect of this improvement
becomes important for thin films. These plots show that the
transitions between stable phases occur at virtually the same
value of $\Delta$ in the OP-NLDF theory and in the improved SFM.
Furthermore, the application of the latter approach for planar
slits also yields better agreement with NLDF results as shown in
Fig.\ \ref{phase4}.

According to our calculations almost all the examined
adsorbate-substrate combinations will form CC at $T=T_t$, the
exceptions are Ne, Ar, Kr, and Xe adsorbed into cavities of Cs. In
these cases it becomes necessary an extra determination of critical
temperatures for CC like it has been done in \cite{sur}. In this
respect, it should be mentioned that it has been previously found
that neither Ne confined by planar walls of Cs \cite{gat0} nor Ne
and Ar adsorbed into cylinders of Cs \cite{sur} form CC at $T =
T_t$, showing a systematic dependence on the pore curvature.

\begin{figure}
\centering\includegraphics[width=7cm]{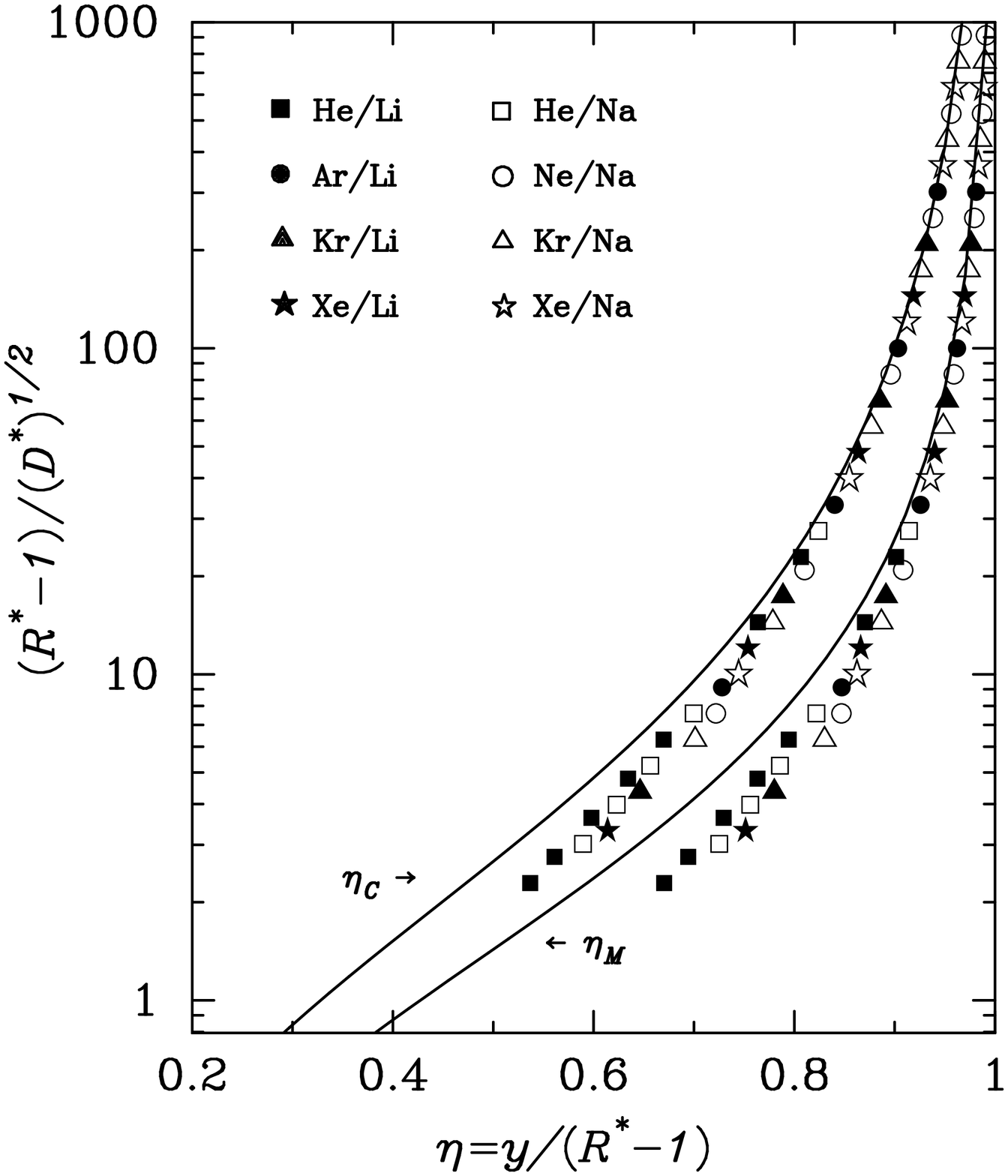}
\caption{Nearly universal correspondence of critical values of
reduced inner radii $\eta_M$ and $\eta_C$ and the dimensionless
ratio of effective radius with potential strength $(R^* -
1)/\protect\sqrt{D^*}$ in the single hysteretic loop. The solid
curves are asymptotic functions given by Eqs.\
(\protect\ref{Psi1}) and (\protect\ref{Psi2}).}
\label{coals}
\end{figure}

The critical radii for metastable phases and unstable films
displayed in Fig.\ \ref{coals} resemble the behavior found
previously for cylindrical pores \cite{sur,coal0}. So, adsorption
potentials built up for different curved geometries lead to
qualitatively similar features. On the other hand, it is suggested
how an elementary hysteretic cycle could be constructed by filling
and draining the cavity along paths determined by the critical
points ($\Delta_M, \eta_M$) and ($\Delta_C, \eta_C$).

In summary, we can state that the present work close successfully
the application of the so called SFM for studying CC in pores
exhibiting standard regular geometries, i.e. planar slits,
cylindrical pores and spherical cavities.

\acknowledgements

We thank Dr. E.S. Hern\'andez for fruitful discussions.
This work was supported in part by the Ministry of Culture and
Education of Argentina through Grants ANPCyT No.
PICT2000-03-08540 and UBACYT No. EX-01/X103.

\newpage

\begin{table}
\caption{Experimental values of relevant observables for the inert
gases in the liquid phase at the triple point and the lattice
parameters of solid alkali metals.}
\begin{tabular}{lccccr}
System & $T_{t}$~[K] & $\mu_0$~[K] & $\rho_0$~[\AA$^{-3}$]
& $\sigma_{lv}$~[K/\AA$^2$] & Ref. \\
\tableline
 \\
$^4$He & \dec 0.$^a$ & -7.15 & 0.02184 && \onlinecite{sat2} \\
       &&         &         & $0.274 \pm 0.003$
                            & \onlinecite{gest,edda} \\
       &&         &         & $0.257 \pm 0.001$
                            & \onlinecite{isis} \\
       &&         &         & $0.272 \pm 0.002$
                            & \onlinecite{raw} \\
Ne  & \dec 24.55  &  -232. & 0.03694 & \dec 3.98
                               & \onlinecite{chen,wy,brush} \\
Ar  & \dec 83.81  &  -930. & 0.02117 & \dec 9.74
                               & \onlinecite{chen,wy,brush} \\
Kr  & \dec 115.76 & -1342. & 0.01785 & \dec 11.22
                                & \onlinecite{chen,wy,brush} \\
Xe  & \dec 161.39 & -1907. & 0.01411 & \dec 12.65
                                & \onlinecite{chen,wy,brush} \\
 \\
       & & $a$~[\AA]$^b$ & $\rho_{\rm sub}$~[\AA$^{-3}$]$^c$ &
       & Ref. \\
\tableline
Cs     & & $6.045$ & $0.009054$ && \onlinecite{kit} \\
Rb     & & $5.585$ & $0.0115 $ && \onlinecite{kit} \\
K      & & $5.225$ & $0.0140 $ && \onlinecite{kit} \\
Na     & & $4.225$ & $0.02652$ && \onlinecite{kit} \\
Li     & & $3.491$ & $0.04701$ && \onlinecite{kit} \\
\end{tabular}
$^a$ The data for $^4$He correspond to $T=0$~K. \\
$^b$ From Ref. \onlinecite{kit}. \\
$^c$ Body centered cubic crystal structure.
\label{table1}
\end{table}

\begin{table}
\caption{Values of the LJ parameters for the interaction between
noble gases and alkali metals, together with the asymptotic well
depth and the location of the minimum corresponding to potentials
given by Eq.\ (\protect\ref{sub1}).}
\begin{tabular}{lccrcr}
System & $\varepsilon_{LJ}$~[K]$^a$ & $\sigma_{LJ}$~[\AA]$^a$
& $r_m$~[\AA] & $D_\infty$~[K] & $D^*$ \\
\tableline
He-Cs & \dec   1.21 & 6.47 & 5.55 & \dec  6.55 &  5.84 \\
He-Na & \dec   1.73 & 5.40 & 4.63 & \dec 15.95 & 11.87 \\
He-Li & \dec   1.92 & 5.22 & 4.48 & \dec 28.34 & 20.39 \\
Ne-Cs & \dec   8.65 & 5.23 & 4.49 & \dec 24.73 &  2.06 \\
Ne-Na & \dec  11.94 & 4.37 & 3.75 & \dec 58.34 &  4.06 \\
Ne-Li & \dec  13.33 & 4.23 & 3.63 & \dec 104.7 &  7.06 \\
Ar-Cs & \dec  51.79 & 4.86 & 4.17 & \dec 118.8 &  2.15 \\
Ar-Na & \dec  60.32 & 4.20 & 3.60 & \dec 261.6 &  4.10 \\
Ar-Li & \dec  66.00 & 4.08 & 3.50 & \dec 465.2 &  7.08 \\
Kr-Cs & \dec  87.16 & 4.75 & 4.08 & \dec 186.7 &  2.42 \\
Kr-Na & \dec  94.10 & 4.16 & 3.57 & \dec 396.6 &  4.51 \\
Kr-Li & \dec 100.10 & 4.06 & 3.48 & \dec 695.2 &  7.71 \\
Xe-Cs & \dec 117.2  & 4.87 & 4.18 & \dec 270.6 &  2.52 \\
Xe-Na & \dec 126.0  & 4.27 & 3.66 & \dec 574.3 &  4.70 \\
Xe-Li & \dec 135.8  & 4.16 & 3.57 & \dec 1015. &  8.08 \\
\end{tabular}
$^a$ Parameters calculated with data taken from \protect\cite{pat}
by applying the procedure outlined in \protect\cite{anil}. \\
\label{table2}
\end{table}

\newpage

\eject

\newpage

\begin{references}
\bibitem[+]{urrucic}Also at the Comisi\'on de Investigaciones
Cient\'{\i}ficas de la Prov. de Buenos Aires, Calle 526 entre 10 y
11, RA--1900 La Plata, Argentina.
\bibitem[*]{szybco}Also at the Carrera del Investigador
Cient\'{\i}fico of the Consejo Nacional de Investigaciones
Cient\'{\i}ficas y T\'ecnicas, Av. Rivadavia 1917,
RA--1033 Buenos Aires, Argentina.
\bibitem{chen}E.\ Cheng, M.\ W.\ Cole, W.\ F.\ Saam, and J.\
Treiner, Phys.\ Rev.\ B {\bf 48}, 18214 (1993).
\bibitem{chiz}A.\ Chizmeshya, M.\ W.\ Cole, and E. Zaremba,
J.\ Low Temp.\ Phys.\ {\bf 110}, 677 (1998).
\bibitem{gat0}S.\ M.\ Gatica, M.\ M,\ Calbi, and M.\ W.\ Cole,
Phys.\ Rev.\ E {\bf 59}, 4484 (1999).
\bibitem{calb}M.\ M,\ Calbi, F.\ Toigo, S.\ M.\ Gatica, and M.\
W.\ Cole, Phys.\ Rev.\ B {\bf 60}, 14935 (1999).
\bibitem{cyan}L.\ Szybisz, Physica A {\bf 283}, 193 (2000). 
\bibitem{less}L.\ Szybisz and S.\ M.\ Gatica, Phys.\ Rev.\ B {\bf
64}, 224523 (2001).
\bibitem{gat1}S.\ M.\ Gatica, M.\ M.\ Calbi, M.\ W.\ Cole, Phys.\
Rev.\ E {\bf 65} 061605 (2002).
\bibitem{sur}L.\ Szybisz and I.\ Urrutia, Phys.\ Rev.\ E {\bf 66},
051201 (2002).
\bibitem{urs}I.\ Urrutia and L.\ Szybisz, Physica A, in press
(2004).
\bibitem{coal0}M.\ W.\ Cole and W,\ F.\ Saam, Phys.\ Rev. Lett.\
{\bf 32}, 985 (1974).
\bibitem{sus}E.\ S.\ Hern\'andez, M.\ W.\ Cole, and M.\
Boninsegni, Phys.\ Rev.\ B {\bf 68} 125418 (2003).
\bibitem{sur1}L.\ Szybisz and I.\ Urrutia, J.\ Low Temp.\ Phys.\
{\bf 134}, 1079 (2004).
\bibitem{bojan99}M.\ J.\ Bojan, G.\ Stan, S.\ Curtarolo, W.\ A.\
Steele, and M.\ W.\ Cole, Phys.\ Rev.\ E {\bf 59}, 864 (1999).
\bibitem{curta00}S.\ Curtarolo, G.\ Stan, M.\ J.\ Bojan, M.\ W.\
Cole, and W.\ A.\ Steele, Phys.\ Rev.\ E {\bf 61}, 1670 (2000).
\bibitem{ancil01}F.\ Ancilotto, S.\ Curtarolo, F.\ Toigo, and
M.\ W.\ Cole,  Phys.\ Rev.\ Lett.\ {\bf 87}, 206103 (2001).
\bibitem{ancil99}F.\ Ancilotto and F.\ Toigo, Phys.\ Rev.\ B
{\bf 60}, 9019 (1999).
\bibitem{hill}T.\ L.\ Hill, Adv.\ Catalysis {\bf 4}, 211 (1952).
\bibitem{ricca}F.\ Ricca, Nuovo Cim.\ Suppl.\ {\bf V}, 339 (1967).
\bibitem{which}I.\ Derycke, J.\ P.\ Vigneron, Ph.\ Lambin, A.\ A.\
Lucas, and E.\ G.\ Derouane, J.\  Chem.\  Phys.\ {\bf 94}, 4620
(1991).
\bibitem{pat}S.\ H.\ Patil, J.\ Chem.\ Phys.\ {\bf 94}, 8089
(1991).
\bibitem{anil}F.\ Ancilotto, E.\ Cheng, M.\ W.\ Cole, and F.\
Toigo, Z.\ Phys.\ B: Condens. Matter {\bf 98}, 323 (1995).
\bibitem{kit}C.\ Kittel, {\em Introduction to Solid State Physics}
(Wiley, New York, 1986), Chap. 1.
\bibitem{sat2}S. Stringari and J. Treiner, Phys.\ Rev.\ B {\bf 36},
8369 (1987).
\bibitem{gest}H.\ M.\ Guo, D.\ O.\ Edwards, R.\ E.\ Sarwinski, and
J.\ T.\ Tough, Phys.\ Rev.\ Lett.\ {\bf 27}, 1259 (1971).
\bibitem{edda}D.\ O.\ Edwards and W.\ F.\ Saam, in {\it Progress in
Low Temperature Physics}, edited by D.\ F.\ Brewer (North-Holland,
Amsterdam, 1978) Vol. 7A, Chap. 4.
\bibitem{isis}M. Iino, M. Suzuki, and A.\ J.\ Ikushima, J.\ Low
Temp.\ Phys.\ {\bf 61}, 155 (1985).
\bibitem{raw}P. Roche, G. Deville, N.\ J.\ Appleyard, and F.\ I.\
B.\ Williams, J.\ Low Temp.\ Phys.\ {\bf 106}, 565 (1997).
\bibitem{wy}S.-T.\ Wu and G.-S.\ Yan, J.\ Chem.\ Phys.\ {\bf 77},
5799 (1982).
\bibitem{brush}L.\ W.\ Bruch, M.\ W.\ Cole, and E.\ Zaremba, {\em
Physical Adsorption} (Oxford Univ. Press, Oxford, 1997).
\bibitem{stucco}G.\ Stan and M.\ W.\ Cole, Surf.\ Sci.\ {\bf 395},
280 (1998).
\bibitem{row}J.\ S.\ Rowlinson and B.\ Widom, {\em Molecular Theory
of Capillarity} (Clarendon Press, Oxford, 1982).
\bibitem{dhpt}J.\ Dupont-Roc, M.\ Himbert, N.\ Pavloff, and J.\
Treiner, J.\ Low Temp.\ Phys.\ {\bf 81}, 31 (1990).
\bibitem{sur2}L.\ Szybisz and I.\ Urrutia,  Phys.\ Rev.\ B
{\bf 68}, 054518 (2003).
\bibitem{chen0}E.\ Cheng, M.\ W.\ Cole, W.\ F.\ Saam, and J.\
Treiner, Phys.\ Rev.\ B {\bf 46}, 13967 (1992).
\bibitem{hall1}M.\ P.\ Lilly and R.\ B.\ Hallock, Phys.\ Rev.\ B
{\bf 63}, 174503 (2001).

\end{references}
\end{document}